# Study of the influence of solar variability on a regional (Indian) climate: 1901-2007

O.P.M. Aslam, Badruddin*

*Department of Physics, Aligarh Muslim University, Aligarh-202002, India.*

**Abstract**

We use Indian temperature data of more than 100 years to study the influence of solar activity on climate. We study the Sun-climate relationship by averaging solar and climate data at various time scales; decadal, solar activity and solar magnetic cycles. We also consider the minimum and maximum values of sunspot number (SSN) during each solar cycle. This parameter SSN is correlated better with Indian temperature when these data are averaged over solar magnetic polarity epochs (SSN maximum to maximum). Our results indicate that the solar variability may still be contributing to ongoing climate change and suggest for more investigations.

**\*Corresponding author. Tel/Fax: +91 571 2701001*
 *Email address: badr.physamu@gmail.com (Badruddin)*

## 1. Introduction

The long-term increase in the globally averaged yearly mean temperatures registered in the 20$^{th}$ century has raised the question as to what part, if any, of the observed changes can be attributed to human influence and what part, if any, can be attributed to natural phenomena? A measure of natural phenomenon, the sunspot number (SSN) is a solar activity index with long data record. It is frequently used when studying long-term phenomena like climate change, though it may not be the most appropriate index (Georgieva and Kirov, 2006).

The term global warming is now popularly used to refer the recent reported increase in the mean surface temperature of the Earth; this increase being attributed to increasing human activity, and in particular to the increased contribution of greenhouse gases (Carbon dioxide, Methane and Nitrous oxide) in the atmosphere. However, there is a dissenting view of global warming science too, which is at odds with this view of the cause of global warming (see, Khandekar et al., 2005). The physical mechanism of the greenhouse gases has been understood whereas the mechanism of solar influence on weather and climate requires more detailed study (Stozhkov, 2003; Gray et al., 2010).

Observations over the last century have shown that the climate at most places on our globe has changed considerably. The extent to which these changes result from human and/or natural forcing is a subject of intense study (e.g., Solanki and Krivova, 2003; Hiremath, 2009; Mufti and Shah, 2011; Rao, 2011; Ahluwalia, 2012). One reason is that both human influences on the environment (e.g., anthropogenic $CO_2$ in the atmosphere) and solar activity increased considerably over the last century. This covariance hampers isolation of their separate effects. Moreover, the climatic impacts of several forcing factors are still insufficiently understood.

The solar-climate relationship is currently a matter of a fierce debate. Despite the increasing evidence of its importance, solar-climate variability is likely to remain controversial until a physical mechanism is established. Nevertheless, it is important to identify the primary forcing agents since they provide the fundamental reason why the climate changed. A key issue of climate change is to identify the forcing and their relative contributions.

The Sun can have obvious effect on climate change; its radiation is the main energy source for the outer envelopes of our planet. Nevertheless, there is a long-standing controversy on whether solar variability can significantly generate climate change, and how this might occur. Eddy (1976) initiated the modern study of this topic by pointing out that the Maunder Minimum (1645-1715) in sunspot activity corresponded to the oldest excursion of the Little Ice Age (1450-1850). Subsequent studies related to the solar influence on the Earth's temperature are quite extensive (see, Eddy, 1976; Reid, 1987; Friis-Christensen and Lassen, 1991; Carslaw et al., 2002; Shaviv and Viezer, 2003; de Jager, 2005; de Jager and Usoskin, 2006; Usoskin and Kovaltsov, 2006; Haigh, 2007; Kirkby, 2007; Gray et al., 2010; Beig, 2011; Singh et al., 2011; Hady, 2013, and references therein), indicating the importance of the problem and that there are many issues that require further investigations. There are indications from recent and past research (e.g., see Lockwood et al., 2011; Maghrabi and Al Dajani, 2014; and references therein) that some regional climates will be more susceptible to solar changes.

## 2. Methodology

In this article, we study the solar influence on climate using a climate (temperature) and a solar activity (sunspot) parameter. We used Indian monthly surface temperature data (all India maximum and minimum) anomalies for the period 1901 - 2007. All India monthly maximum (Tmax) and minimum (Tmin) surface temperature data sets for the period 1901 - 2007 are available through Indian Institute of Tropical Metrology's (IITM's) data archival (http://www.tropmet.res.in/). This data archival contains data of 107 years (each year have 12 monthly values), generated at IITM using instrumental meteorological records of the India Meteorological Department (IMD). They used climatological normals of monthly mean maximum and minimum temperatures for the period 1951 - 80 for 388 well-spread stations from the monthly weather records of the India Meteorological Department (IMD, 1999). The procedure adopted for monthly all India maximum and minimum temperature data generation has been explained (see, Kothawale and Rupa Kumar, 2005; ftp://103.251.184.5/pub/data/txtn/README.pdf). We have calculated the anomalies in all India maximum (dTmax) and minimum (dTmin) temperature using this monthly temperature data. We have calculated the mean-yearly temperature by taking average over 12 months (January - December). Average of yearly-mean temperature for the period 1951 – 80 is taken as reference; we calculated the deviation (anomaly in temperature) in yearly-mean temperature of individual years. Using the anomalies in all India maximum and minimum temperature we also calculated anomalies in average temperature [i.e., dTav = (dTmax + dTmin)/2].

Sunspot number is the solar activity parameter available and well documented on monthly and yearly average basis for continuous long periods of time (http://solarscience.msfc.nasa.gov/). The relationship between the anomalies in Indian temperature and SSN has been studied at various time scales relevant for extracting some physical meaning to the Sun-climate relationship.

## 3. Results and discussion

Fig. 1 shows the sunspot variations from the beginning of the last century. To start with, we have studied the relationship between the decadal averages of sunspot number (<SSN>) and temperature anomalies (<dTmax>, <dTmin>, and <dTav> (see, Fig. 2a). We observe that, averaged over decadal time scale, there is some correspondence between the solar and climate parameters. However, this correspondence is poor during the last decade of the 20$^{th}$ century and the beginning years (2001 - 2007) of this century.

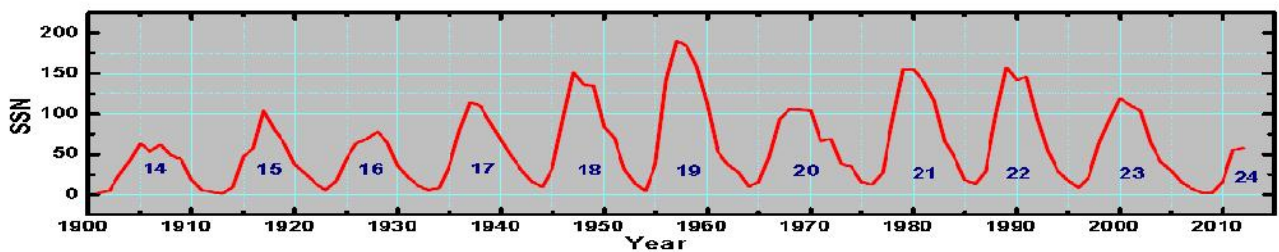

**Fig. 1.** Variation of the annual average sunspot numbers (SSN) from 1901 to 2012.

A Schwabe (solar activity) cycle period may range from 9 to 17 years, measured from one sunspot minimum to the next sunspot minimum, with an average period of 11 years. The Schwabe cycle is the most prominent periodicity in solar activity. It may be more useful to consider the solar cycle average, instead of decadal average, for the study of the relationship between solar activity and climate. Both the durations and the amplitudes of different solar cycles are quite variable (see Fig. 1). Therefore, we performed an analysis to study the Sun-climate relationship for solar cycle averages. We find that relationship between the solar cycle averaged sunspot number (<SSN>) and temperature anomalies is not much different from that observed on the decadal average scale (see, Table 1). Results of correlation analysis between SSN and temperature anomalies at various time scales are tabulated in Table 1, showing value of correlation coefficients ($R$) along with p-value and confidence interval for confidence level of 95%. The probability of error that is involved in accepting our observed result is represented by p-value, *i.e.,* smaller the p-value, stronger the validity of the observed result (*e.g.,* a p-value of 0.05 indicate that there is a 5% probability that the relation between parameters found in our study is a chance of occurrence). Confidence interval provides a range of possible $R$ values which is likely to include an unknown population. Confidence interval includes zero means the correlation is not significant at the given level of confidence (95%).

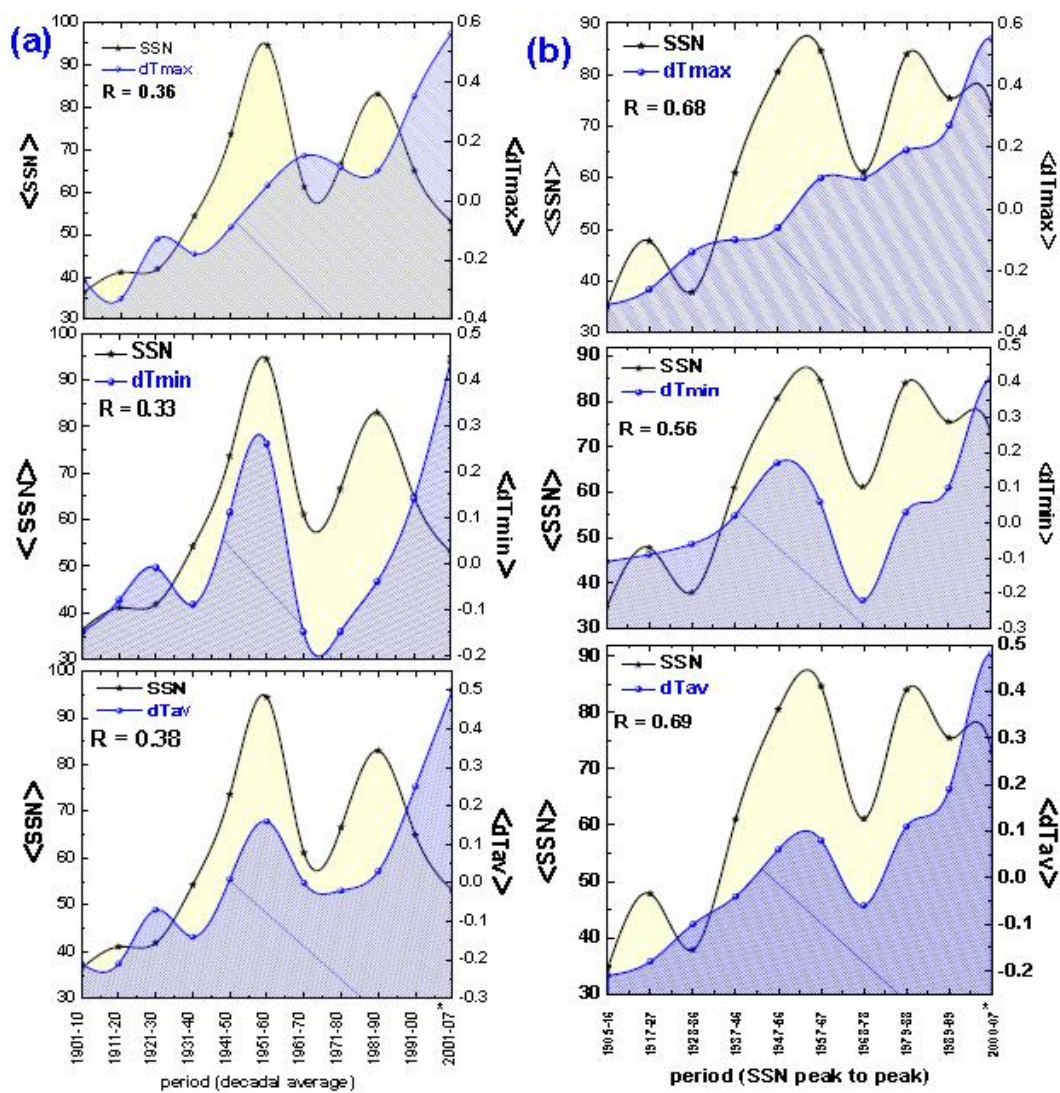

**Fig. 2.** Variation of the SSN and temperature anomalies of Indian temperature for (a) 'decadal average' for the period 1901-2007, (b) the 'SSN Peak to peak averages' for the period 1905-2007.

Reversal in the polarity of the solar polar magnetic field takes place near the solar activity maximum in each solar cycle, and the large-scale interplanetary magnetic field is an extension of the solar polar magnetic field in space (Smith et al., 1978). It is also known that the large-scale structure of the interplanetary magnetic field is of basic importance for the long-term modulation of galactic cosmic rays (Venkatesan and Badruddin, 1990; Kudela et al., 2000; Badruddin et al., 2007). There are indications that long-term variability in cosmic ray intensity influences the Earth's climate (Svensmark and Friis-Christensen, 1997; Kirkby, 2007; Rao, 2011). Thus, we have studied the Sun-climate relationship by averaging the data over the time scales of solar polarity epoch (peak to peak SSN). Averaged over this time scale, we found a significant improvement in correlation between <SSN> and temperature anomalies as compared to decadal and solar activity cycle timescales (see Table 1 and Fig. 2b).

**Table 1**

Result of correlation analysis between sunspot number (SSN) and temperature anomalies at various time scales.

| Time scale | | | Correlation coefficient | | | | | | | | | | | |
|---|---|---|---|---|---|---|---|---|---|---|---|---|---|---|
| | | | dTmax | | | | dTmin | | | | dTav | | | |
| SSN | Temperature | Total periods | R | P-value | Confidence interval of R for confidence level of 95% | | R | P-value | Confidence interval of R for confidence level of 95% | | R | P-value | Confidence interval of R for confidence level of 95% | |
| | | | | | Lower limit | Upper limit | | | Lower limit | Upper limit | | | Lower limit | Upper limit |
| Decadal average | Same period average | 1901-2000 | 0.61 | 0.0611 | -0.032 | 0.896 | 0.63 | 0.0509 | 0.001 | 0.902 | 0.74 | 0.0144 | 0.207 | 0.934 |
| | | 1901-2007* | 0.36 | 0.2768 | -0.306 | 0.789 | 0.33 | 0.3216 | -0.336 | 0.776 | 0.38 | 0.2490 | -0.285 | 0.798 |
| Solar cycle average | Same period average | 1901-1995 | 0.49 | 0.1806 | -0.256 | 0.871 | 0.53 | 0.1422 | -0.207 | 0.883 | 0.75 | 0.0199 | 0.171 | 0.944 |
| | | 1901-2007* | 0.30 | 0.3997 | -0.406 | 0.782 | 0.29 | 0.4163 | -0.415 | 0.778 | 0.35 | 0.3215 | -0.359 | 0.803 |
| SSN peak to peak average | Same period average | 1905-1999 | 0.79 | 0.0113 | 0.265 | 0.954 | 0.67 | 0.0483 | 0.011 | 0.923 | 0.89 | 0.0013 | 0.552 | 0.977 |
| | | 1905-2007* | 0.68 | 0.0305 | 0.088 | 0.917 | 0.56 | 0.0923 | -0.108 | 0.880 | 0.69 | 0.0272 | 0.107 | 0.920 |
| Max SSN in solar cycle | Solar cycle average | 1901-1995 | 0.39 | 0.2994 | -0.370 | 0.837 | 0.65 | 0.0581 | -0.025 | 0.918 | 0.72 | 0.0287 | 0.107 | 0.936 |
| | | 1901-2007* | 0.26 | 0.4163 | -0.415 | 0.778 | 0.41 | 0.2393 | -0.296 | 0.826 | 0.38 | 0.2787 | -0.328 | 0.815 |
| Minimum SSN at the beginning of the cycle | Solar cycle average | 1901-1995 | 0.90 | 0.0009 | 0.586 | 0.979 | -0.03 | 0.9389 | -0.681 | 0.647 | 0.80 | 0.0096 | 0.290 | 0.956 |
| | | 1901-2007* | 0.73 | 0.0165 | 0.186 | 0.931 | 0.05 | 0.8909 | -0.598 | 0.659 | 0.54 | 0.1071 | -0.136 | 0.873 |
| Solar magnetic cycle average | Average of same period | 1905-1999 | 0.84 | 0.1600 | -0.163 | 0.989 | 0.64 | 0.0634 | -0.042 | 0.915 | 0.88 | 0.1200 | -0.010 | 0.992 |
| | | 1905-2007* | 0.68 | 0.2066 | -0.506 | 0.976 | 0.53 | 0.1151 | -0.150 | 0.869 | 0.47 | 0.4244 | -0.704 | 0.956 |

\* Including data points of recent period (available up to 2007)

Georgieva et al. (2012) recently studied the influence of solar poloidal and solar toroidal-related solar activities on the atmospheric circulation. The highest value of sunspots in a solar cycle, $SSN_{max}$, is considered as a proxy for the toroidal field strength (de Jager, 2005). Since the amplitudes of solar activity ($SSN_{max}$) in different cycles are quite variable, ranging from ~50 to ~200 (see Fig. 1). Moreover, the lowest activity ($SSN_{min}$) in the beginning of each cycle is also somewhat variable; we also looked at the relationship between $SSN_{max}$ of each solar cycle and cycle-averages of temperature anomalies, as well as between $SSN_{min}$ at the beginning of each cycle and cycle averages of temperature anomalies. The correlations in these two cases are, in general, lower than those found when averaged over peak to peak sunspot periods (see Table 1).

It is well known that the number of sunspots increases and then decreases in approximately 11-year intervals. The 11-year sunspot cycle is actually a 22-year cycle in the solar magnetic field, with sunspots showing the same hemispheric magnetic polarity on alternate 11-year cycles; polarity reversal taking place around solar maximum. Therefore, we have looked for the relationship between the SSN and the temperature anomalies averaged over the solar magnetic cycles. In this case the correlation of <SSN> with temperature anomalies is somewhat lower as compared to that when averaged over only one polarity epoch. Thus, when averaged over each solar polarity epoch (sunspot maximum to maximum), the relationship between the sunspot number and temperature anomaly is found to be the best among all those discussed above.

The question of a definite relation between temperature and the solar activity is still a matter of debate. Our results, however, indicate that some relationship does exist.

Sunspot cycle 23 was unusual (e.g., see Aslam and Badruddin, 2012; Hady, 2013, and references therein), the current sunspot minimum has been unusually long, and with more than 670 days without sunspots through June 2009. The solar wind is reported to be in a unique low energy state since space measurements began nearly 40 years ago (Fisk and Zhao, 2009; Livingston and Penn, 2009). Unfortunately, all India temperature data ($T_{max}$, $T_{min}$) is not available to us after 2007; it would be very interesting to look for Sun-climate relationship during cycle 24.

## 4. Conclusions

Comparison of the relationships between the Indian temperature anomalies and solar activity (SSN) provides evidence favouring a mechanism that depends not only on the level of sunspot activity but also on solar polarity. In spite of the evidences found during the most part of the past century, the latest temperature rise in the 1990's (especially during solar cycle 23) is difficult to comprehend from most of the discussed results. However, on the solar polarity scale (sunspot maximum to maximum), there are some indications, from the data up to 2007, that a link between solar activity and climate can still be accounted for, and this connection will be watched with curiosity during the current solar cycle 24 and later periods.

**Acknowledgements**

We thank Indian Institute of Tropical Meteorology, Pune and India Meteorological Department for the availability of Indian Climate data. Use of sunspot data through the NGDC is acknowledged. We also thank the Editor and Reviewers, whose comments and suggestions helped us to improve the paper.